\documentstyle[aps,prl]{revtex}

\begin {document}

  \title {Selection Rules for Transport Excitation Spectroscopy of
    Few-Electron Quantum Dots}

  \author {Daniela Pfannkuche}
  \address {Max-Planck-Institut f\"ur Festk\"orperforschung,
            Heisenbergstr. 1,
            D-70569 Stuttgart, Germany}
  \author {Sergio E. Ulloa}
  \address {Department of Physics and Astronomy and Condensed Matter and
            Surface Sciences Program, Ohio University,
            Athens, OH 45701-2979}
  \date{\today} 

 \twocolumn
  \maketitle

  \begin {abstract}
    Tunneling of electrons traversing a few-electron quantum dot is strongly
    influenced by the Coulomb interaction leading to Coulomb blockade effects
    and single-electron tunneling. We present calculations which demonstrate
    that correlations between the electrons cause a strong suppression of most
    of the energetically allowed tunneling processes involving excited dot
    states. 
    The excitation of center-of-mass modes, in contrast,  is unaffected by
    the Coulomb interaction. Therefore, channels connected to these modes
    dominate the excitation spectra in transport measurements.
  \end   {abstract}


  \narrowtext

Transport measurements  are a highly sensitive tool for the investigation of
the electronic structure of semiconductor quantum dots. While 
the differential conductance of such a system in the linear transport
regime is dominated by the classical Coulomb blockade effect
\cite{Houten92:167}, quantum mechanics leaves its fingerprints in magnetic
field dependent addition spectra
\cite{McEuen91:1926,Ashoori93:613} and in excitation spectra
which can be obtained from conductance measurements in the finite drain-source
voltage regime \cite{Johnson92:1592,Weis93:4019,Foxman93:10020}.

Due to a small electronic effective mass, quantum dots in semiconductor
heterostructures exhibit 
a discrete level spectrum, easily discernible at low temperatures. 
Since the charge of these quantum dots can be controlled
at the single-electron level, they are often called {\em artificial atoms}.
The lateral confinement potential which binds 
the electrons to the quantum dot is typically created by spatially
extended charge distributions. Therefore, it obeys a parabolic
dependence on the distance from the center of the system ($\propto
r^2$) rather than the $1/r$-dependence characteristic of the
core potential of {\em natural} atoms \cite{Kumar90:5166}. 
In analogy to the optical spectroscopy on conventional atoms the first
attempts  to study the excitation spectrum of these artificial atoms have been
made by far-infrared (FIR) spectroscopy
\cite{Sikorski89:2164,Demel90:788,Lorke90:2559,Meurer92:1371}.
However, the long wavelength of the far-infrared radiation
together
with the parabolic confinement potential  prohibits most transitions into
excited states by strict dipole selection rules
\cite{Maksym90:108,Li91:5151}.

Transport spectroscopy on the other hand is expected not to suffer from
these restrictions, but to allow a unique full spectroscopy of the level
structure 
in these artificial atoms. Nevertheless, experiments on few-electron
quantum dots exhibit only a sparse excitation spectrum with relatively
large level spacings \cite{Weis93:4019}. These measurements suggest that the
dominant resonances are due to a constant level spacing, independent of the
number of electrons in the quantum dot. This behavior,
which could be understood for a non-interacting electron system, is
in\-deed remarkable, since the Coulomb interaction strongly influences
the energy spectrum of the quantum dot and leads to small level
spacings and strong correlations \cite{Maksym90:108,Pfannkuche93:2244}.

Thus, the experimentally observed large level spacing together with
the appearance of a characteristic excitation energy demands an
explanation. We present calculations which show that even in transport
experiments the excitation
spectra are dominated by center-of-mass modes, similar to the situation in FIR
spectroscopy. This gives rise to the constant level 
spacing with the characteristic single-particle excitation energy. This
situation is caused by a suppression of most other transitions due to strong
{\em correlations} between the electrons.

We study the transport properties of a two-dimensional quantum dot coupled to
two reservoirs by tunneling barriers in the conventional tunneling Hamiltonian
approach \cite{Kinaret92:4681}. The Hamiltonian of the system is given by a
sum of the dot Hamiltonian $H_D$, a reservoir Hamiltonian $H_R$, and a
tunneling term $H_T$. The tunneling Hamiltonian describes the transfer of
electrons from the reservoir, where the Coulomb interaction between the
electrons is effectively screened (metallic regime), to the quantum dot, where
interactions are most important.
Our results are based on features of the few-particle eigenstates
of the quantum dot which we obtain from a numerical diagonalization 
of the dot Hamilton operator
\begin {equation}
  H_D = \sum_n \epsilon_n d_n^+ d_n
        + \sum_{n,m,n',m'} V_{n m  n' m'}
          d_n^+ d_m^+ d_{n'} d_{m'}.
  \label{H_dot}
\end   {equation}
It describes interacting electrons in a parabolic confinement
potential 
subjected to a perpendicular magnetic field $B$. The single-particle energies
$\epsilon_n$ are connected with left and right circularly polarized
oscillator eigenmodes,
\begin {equation}
   \epsilon_n =  \hbar \Omega_+ \left(N_+ + 1/2 \right)
               + \hbar \Omega_- \left(N_- + 1/2 \right)
  \label{H_CM}
\end   {equation}
with $\Omega_\pm = (\sqrt{4\Omega_0^2 + \omega_c^2} \pm \omega_c)/2$.
$\Omega_0$ characterizes the confining potential and $\omega_c = e B / (m c)$
is the cyclotron frequency. A crucial feature of the 
parabolic confinement potential is the separation of the dot Hamiltonian into a
relative and a center-of-mass (CM) part \cite{Wagner92:1951,Hawrylak93:485},
where the spectrum of the CM-Hamiltonian is identical to the single-particle
spectrum (\ref{H_CM}) and independent of the number of electrons in the dot.
It is this separation which prohibits in FIR~spectroscopy the observation of
the spectral fine structure of these 
artificial atoms, and makes visible only the CM-excitations.
The Coulomb interaction between electrons enters only the
Hamiltonian of the internal degrees of freedom (relative part).
We obtain the matrix elements for the interaction, 
$V_{n m n' m'}$, in closed form for the usual $1/(\varepsilon
r)$ Coulomb potential with the dielectric constant $\varepsilon$.

Regarding only sequential tunneling, the current through the quantum
dot is given by \cite{Kinaret92:4681}
\begin {eqnarray}
  I = -e \sum_{\alpha,\alpha'} & \hspace {-.5cm}\Gamma_{\alpha \alpha'}\;  
  \big[ P(N,\alpha)+P(N-1,\alpha') \big] \hfill \nonumber \\
  & \times  
  \big[ f(\Delta E_{\alpha \alpha'} -\mu_l) - 
         f(\Delta E_{\alpha \alpha'} -\mu_r) \big]
  \label {current}
\end   {eqnarray}
where the resonant energy $\Delta E_{\alpha \alpha'} = E(N,\alpha) -
E(N-1,\alpha')$ is the 
difference between the energy of a $N$-particle state $\alpha$ and a
$(N-1)$-particle state $\alpha'$. The
Fermi-Dirac distribution function $f$ characterizes the occupation of electron
levels in the left (electro-chemical potential $\mu_l$) and right ($\mu_r$)
reservoir.
The probability $P(N,\alpha)$ to find the quantum dot in the $N$-particle
state $\alpha$ will deviate from its equilibrium value for a given
drain-source voltage $(\mu_l - \mu_r)/e$. Its dependence on the tunneling rate
$\Gamma_{\alpha \alpha'}$ is well described by kinetic equations
\cite{Kinaret92:4681,Beenakker91:1646} and leads to the blocking
of conducting channels and negative differential conductance
\cite{Weis93:4019,Kinaret92:4681,Weinmann94:467}.

In a ``random phase'' approximation, i.e. neglecting phase correlations between
the initial state of the electron in the lead and its final state in the
quantum dot, the tunneling rate $\Gamma_{\alpha \alpha'}$ factorizes into an
effective tunneling rate for a non-interacting electron traversing 
the barrier,  $\gamma$, and an overlap or spectral weight matrix element.
\cite{Kinaret92:4681} (For simplicity, we assume a constant effective
tunneling rate and neglect the dependence of $\gamma$ on the single-electron
dot states.) The overlap matrix element is given by
\begin {equation}
  \sum_n |\langle N,\alpha|\;d_n^+\;|(N-1),\alpha'\rangle|^2.
  \label {overlap}
\end   {equation}
This quantity describes to what extent the compound state, built by an
incoming electron and the $(N-1)$-electron state $\alpha'$ in the dot,
overlaps with the $N$-electron dot state $\alpha$. While for an
uncorrelated electron system this overlap evaluates to unity,
correlations reduce it considerably
\cite{Kinaret92:4681,Palacios93:495}. Due to the summation over all
single-particle dot states with equal weight in eq.~(\ref{overlap})
this quantity is insensitive to any features of the incoming electron.
It only reflects the correlations in the states $\alpha$ and
$\alpha'$.  In the following we will concentrate on these overlap
matrix elements, as they predominantly determine the transport through
the quantum dot (for levels separated by more than $kT$, as discussed for
example in Ref.\ \cite{Kinaret92:4681}.  In experiments, $kT \approx 50$mK
$\approx 4\mu$eV).  
Our numerical results are obtained for quantum dots with up to $N = 3$
electrons.

Fig.~\ref{spectra} shows the spectra of a quantum dot occupied by one,
two or three electrons (QD-hydrogen, QD-helium, or QD-lithium) at zero
magnetic field. Due to the Coulomb interaction the level spacing of the
few-particle spectra is much smaller than the single-electron level
spacing, which is given by the confining energy $\hbar \Omega_0 =
2$~meV
(this value is characteristic of experimental geometries, such as the
system in Ref.\ \cite{Weis93:4019}).
Enforced by energy conservation, tunneling through the quantum dot is
only allowed  when the energy of the incoming particle matches the
energy difference between a QD-helium (lithium) eigenstate and a
QD-hydrogen (helium) eigenstate. The occurrence of all energetically
allowed transitions would clearly result in a dense excitation
spectrum.  That contradicts recent experiments \cite{Weis93:4019}.
Thus, it is not only the energy requirement which determines whether an
electron tunnels through the quantum dot. A mechanism to select certain
transitions is provided by the overlap matrix elements,
eq.~(\ref{overlap}).

One of these selection rules is evident from spin conservation
\cite{Weinmann94:467}. For our few-electron quantum dots they forbid
transitions between a spin singlet state of QD-helium and a spin
quartet state of QD-lithium. This has already consequences for the
usual Coulomb blockade oscillations, as transitions between the ground
states of QD-helium and QD-lithium (G-G-transition) in a
perpendicularly applied magnetic field are considered.  The total spin
of the ground state of these few-electron systems oscillates between
its maximum (triplet, respectively quartet) and its minimum value
(singlet, respectively doublet) when the magnetic field is increased
\cite{Wagner92:1951,Hawrylak93:485}.  Fig.~\ref{ground} shows the
ground state energy of QD-helium and QD-lithium as a function of the
magnetic field $B$. The total spin is indicated for the different field
ranges. For most values of the magnetic field G-G-tran\-si\-tions are
possible. However, as indicated in the figure, two field ranges occur
where the ground state of QD-helium is a singlet while that of
QD-lithium is a quartet: the G-G-transition is blocked. Thus, a drastic
reduction of the conductance in the linear response regime should be
observed when the magnetic field is tuned into these ranges.

However, spin selection rules alone are not sufficient to explain the
sparseness of the observed excitation spectra. Apart from an angular
momentum conservation in the dot, the effect of which is annihilated by
connecting the dot laterally to the leads, the overlap matrix elements,
eq.~(\ref{overlap}), provide no further strict selection rules.
Nevertheless, very small spectral weights strongly reduce the
probability for the occurrence of the corresponding transitions. This
gives rise to {\em quasi-selection rules}. Fig.~\ref{2to3} shows the
values of the overlap matrix elements for transitions from the
QD-helium ground state to all states of QD-lithium at $B=0$. They
correspond to the most important processes, if the quantum dot is in
its ground state before the subsequent tunneling process. Since the
2-particle ground state is a singlet state, only transitions into
doublet states are possible.  Transitions into degenerate states are
subsumed under the same peak, since all corresponding overlap elements
contribute accordingly to the tunneling rate at a given energy $\Delta
E$. This applies especially to the double degeneracy with respect to
the z-component of the doublet spin, giving rise to a factor of two in
the peak heights. At low energies, below 15~meV, the figure exhibits a
few regularly-spaced dominant peaks with several relatively smaller
ones between them. This energy regime coincides with the experimental
situation where only one electron tunnels at a time
\cite{Weis93:4019}.

The first peak at 9.7~meV corresponds to the G-G-tran\-si\-tion. The
$B=0$ states are, apart from the spin degeneracy, degenerate with
respect to the sign of the angular momentum.  Therefore, the 3-particle
ground state (with an orbital angular momentum of $\hbar$) is four-fold
degenerate. Each of these states contributes equally to the peak shown
in Fig.~\ref{2to3}. Thus, the individual overlap matrix elements
(eq.~(\ref{overlap})) are much smaller than unity, indicating the
strong correlations in the few-particle states.  Nevertheless, compared
with the other spectral weights the individual overlap matrix elements
of the G-G-transition are the largest. Therefore, these transitions
occur with the highest probability.
They are followed by a transition which is subsumed in the peak
appearing at 11.7~meV, i.e., at an excitation energy of 2~meV $=
\hbar\Omega_0$ above the G-G-transition. This transition corresponds to
a state of the 3-particle system with one quantum of CM-excitation in
addition to the ground state energy. As a consequence of the separation
of the CM- and relative motion, each transition within the internal
degrees of freedom is accompanied by independent excitations of
CM-modes.  Since the excitation spectrum of the CM-motion is
equidistant, these transitions give rise to equidistant conductance
peak replicas (see arrows in Fig.~\ref{2to3}).  Moreover, at zero
magnetic field the CM-energy spectrum is highly degenerate.  Therefore,
the overlap matrix elements of several transitions contribute to the
tunneling probability for a given resonant energy. It is this
prevalence of the CM-excitations which provides an explanation for the
experimental observation of a characteristic level spacing independent
of the number of particles in the quantum dot:  The CM-spectrum is
independent of the number of particles in a parabolic quantum dot and
equals the single-electron spectrum
\cite{Maksym90:108}.  
[Notice that the small thermal broadening of the Fermi level in
experiments, typically 100 mK, will not blur the different
CM-excitations, occurring with much larger spacings, 2 meV here.]

It is clear that in addition to the G-G-transition, tunneling via other
internal excitation states is possible, with their corresponding
CM-excitation replicas (see second pair of replicas in
Fig.~\ref{2to3}). In some cases (e.g. in the transition from the
single-particle ground state to the first-excited 
2-particle state, not shown here) the overlap matrix elements can even
be larger than those of the G-G-transition. Due to their large
degeneracy, transitions which end in a maximum-spin state produce very
pronounced series of replicas, much stronger than the ones shown in
Fig.~\ref{2to3}. The occurrence of more than one group of dominant
replicas gives rise to smaller level spacings in the spectra. Both
effects have been observed in experiment for different $N$-states in
the dot. However, we found in all cases that only few groups of
dominant replica occurred, while most excitations of internal degrees
of freedom {\em are suppressed in tunneling}.  In our model, with a
constant tunneling rate $\gamma$, this suppression is solely due to
correlations of the few-particle wave functions.  These strong built-in
correlation effects lead to nearly-forbidden transitions.  Notice that
the effective suppression of internally-excited states becomes less
pronounced in the high energy range, as indicated by a more even
distribution of high-overlap peaks in Fig.~\ref{2to3}.

Although our calculations only include systems with up to three
electrons, general conclusions can be drawn for quantum dots occupied
by more electrons.  As long as correlations strongly influence the low
energy excitations, a strong reduction of most overlap matrix elements
occurs, leading to quasi-selection rules.  These will suppress most
transitions involving excitations of the internal degrees of freedom in
the rather dense spectrum.  Since the CM-motion remains unaffected,
these degrees of freedom can be excited easily. Thus, it can be
expected that they would likewise dominate the transport resonances.
Another aspect enhancing the effect of dominant CM-excitations may
arise especially in large quantum dots laterally connected to the
reservoirs by split gate tunnel barriers. This geometry forces the
electron to enter at a certain point at the edge of the dot. This
introduces a large dynamical
dipole moment in the system built out of the incoming electron and the
$(N-1)$-particle state in the dot. As known from far-infrared
investigations, CM-excitations give rise to a strong dipole moment.
Therefore, one would expect that the overlap of such compound state
with a CM-excitation will be larger than others, thus favoring it for
tunneling.

A basic assumption for the discussion so far has been the strict
separation of CM- and relative motion which is only fulfilled in a
truly parabolic confinement potential. However, calculations which take
into account anharmonicities of the confining potential
\cite{Pfannkuche91:13132} have shown that for realistic conditions
\cite{Lier93:14416} the coupling between the CM-motion and the relative
motion is weak, leaving the demonstrated mechanism for dominant
CM-excitations essentially unchanged.

In summary, the analysis of the overlap matrix elements which govern
the tunneling rate for an electron traveling through a quantum dot
shows that the strong correlations present in few-electron dot states
are extremely important. They strongly reduce the probability of
tunneling through channels involving excitations of the internal
degrees of freedom. This leads to a dominance of the center-of-mass
excitations which are not affected by correlation effects. The
center-of-mass excitation spectrum is identical with the
single-particle spectrum of an electron in the parabolic quantum dot,
independent of the number of particles. These features give an
explanation for the experimental observations in the non-linear,
single-electron tunneling regime, showing relatively sparse excitation
spectra with the characteristic single-electron level spacing.

  We would like to thank R. Blick, J. Weis, G. Sch\"on, R. R.
  Gerhardts, and E.  Zaremba for helpful and lively discussions. This
  work has been supported by the Bun\-des\-mi\-nister f\"ur Forschung
  und Technologie, and the US Department of Energy Grant
  No.\ DE--FG02--91ER45334.

\begin {figure}
  \caption {Level scheme of a parabolic quantum dot with 1, 2, or 3 electrons
    at $B = 0$~T. Arrows represent the corresponding addition energies.
    Parameters for all figures: Confining energy $\Omega_0 = 2$~meV, GaAs
    parameters: dielectric  
    constant $\varepsilon = 12.4$, effective mass $m^* = 0.067m_e$.
    } 
  \label{spectra}
\end   {figure}

\begin {figure}
  \caption{Ground state energy of a parabolic quantum dot with 2 and with 3
    electrons as a function of the magnetic field $B$. Spin
    states are indicated: S = Singlet, T = Triplet, D = Doublet, Q = Quartet.
    Arrows mark those field ranges where G-G-transitions are forbidden by spin
    selection rules.
    }
  \label{ground}
\end   {figure}

\begin {figure}
  \caption {Summed values of the overlap matrix elements for transitions
    between the 2-particle 
    ground state in the quantum dot at $B=0$ and all states
    of the 3-electron system as a function of the transition
    energy $\Delta E$. Arrows indicate center-of-mass
    excitations.
    }
  \label{2to3}
\end   {figure}

\end   {document}